\RequirePackage{fix-cm}
\documentclass{iopjournal}

\usepackage{amsmath,amssymb}
\usepackage{listings}
\usepackage{bm}
\usepackage{aas_macros}
\usepackage{tikz}
\usepackage{booktabs}
\usepackage{subcaption}
\usetikzlibrary{arrows.meta, positioning, shapes.geometric, fit}
\usepackage{soul} 
\usepackage[normalem]{ulem} 

\graphicspath{{./images/}}

\definecolor{Cerulean}{RGB}{0,123,167}

\begin{document}

\pagestyle{plain}

\title{Finding black hole spins efficiently during a numerical binary evolution}

\author{Himanshu Chaudhary$^{1,*}$, Rob Owen$^2$, Mark A. Scheel$^1$ and Saul A. Teukolsky$^{1,3}$}

\affil{$^1$Theoretical Astrophysics 350-17, California Institute of Technology, Pasadena, CA, 91125, USA}
\affil{$^2$Department of Physics and Astronomy, Oberlin College}
\affil{$^3$Cornell Center for Astrophysics and Planetary Science, Cornell University, Ithaca, New York 14853, USA}
\affil{$^*$Author to whom any correspondence should be addressed.}

\email{hchaudha@caltech.edu}


\begin{abstract}

The dynamics of a binary black hole system depend on its masses and spins.
For a binary at finite separation, it is not possible to define these quantities in an unambiguous way; however, there are several reasonable definitions that reduce to the expected values in the limit of infinite separation.
Approximate Killing vector (AKV) spin is one of the spin definitions used in the numerical relativity code SpEC.
AKV spin requires finding approximate Killing vectors on an apparent horizon, which reduces to a generalized eigenvalue problem of size $\mathcal{O}(L^2)$, and a direct solve has time complexity $\mathcal{O}(L^6)$,
where $L$ is the highest spherical harmonic mode used to represent the apparent horizon.
This scaling means that the cost of computing AKV spins increases rapidly as we simulate systems at higher resolutions, especially those with high spin or mass ratios. We describe a new algorithm for computing AKV spins that is much faster than the current algorithm.

\end{abstract}

\section{\label{sec:Introduction} Introduction}

Detectors in the LIGO-Virgo-KAGRA (LVK) \cite{sxs25_5,sxs25_6} collaboration require accurate gravitational-wave models to detect and analyze signals from binary black hole (BBH) mergers \cite{GWTC4.0,sxs25_2,sxs25_1,sxs25_3,sxs25_4}. The only first-principles way to produce gravitational-wave models for the full inspiral, merger, and ringdown is numerical relativity (NR) \cite{NR_Baumgarte,NR_shibata,NR_Alcubierre}, which solves Einstein’s equations directly on a computer. NR simulations are very expensive and slow, which limits their direct use for data analysis. Instead, NR simulations are used as a source of ground truth to construct and calibrate other models like surrogate models \cite{sxs25_68,sxs25_71,sxs25_72,sxs25_69,sxs25_70} and Effective-One-Body (EOB) models \cite{sxs25_43,sxs25_44,sxs25_45,sxs25_46,sxs25_49,sxs25_50}, which are then used for detection and data analysis. Until now, the accuracy of Numerical Relativity (NR) simulations has been sufficient for the gravitational-wave detections made by the LVK collaboration. However, with the new generation of gravitational-wave detectors coming online \cite{sxs25_74,sxs25_77,sxs25_78}, many NR simulations will need to be redone at higher resolution to keep the NR error negligible in data analysis \cite{sxs25_83}.

In NR, the masses and spins of black holes are not input
parameters, but instead they are time-dependent output parameters
that are measured from the numerical solution of Einstein's
equations. In practice, achieving some chosen masses and spins even
at the initial time of the simulation involves guessing ``bare''
masses and spins, solving the Einstein constraint equations,
measuring the masses and spins, adjusting the bare values, and
iterating until the measured values match the desired
ones \cite{sxs25_143}.

When running the code SpEC \cite{SpECwebsite} for binary black hole
(BBH) simulations at high resolution, we noticed a dramatic
slowdown that we ultimately traced to the algorithm that computes
the spin of each dynamical black hole. This algorithm, which
computes the approximate Killing vector (AKV) spin \cite{lovelace_binary-black-hole_2008, Owen:2017yaj}, involves solving a
generalized eigenvalue problem on an apparent horizon.  The apparent
horizon is represented as a surface of angularly dependent
coordinate radius $r_{\mathrm{AH}}(\theta,\phi)=\sum_{\ell,m} S_{\ell m}
Y_{\ell m}(\theta,\phi)$, where $Y_{\ell m}$ are spherical harmonics,
$S_{\ell m}$ are coefficients,
and the sum goes up to some value $L$ that depends on the
resolution.  The reason for the slowdown is that the AKV eigenvalue
problem involves a matrix of size $L^2 \times L^2$, and
for high resolution, $L$ can grow to 100 or more, especially for
large spins or extreme mass ratios.  The AKV calculation is performed on a single core, since it was (incorrectly) assumed to be a
small computational cost compared to the remainder of the BBH
evolution.

The outline of this paper is as follows. Section \ref{sec:AKV spin definition} has some mathematical background about how the AKV spins are defined. Then, in Section \ref{sec:Current AKV algorithm}, we briefly describe the existing algorithm and why it slows down rapidly as the systems being evolved get complicated. In Section \ref{sec:Sparse AKV algorithm}, we describe the new algorithm, which is much faster. Finally, we discuss possible future improvements in Section \ref{sec:Future improvements} and present conclusions in Section \ref{sec:Conclusion}.

\section{\label{sec:AKV spin definition} AKV spin definition}
In this section, we only briefly describe the algorithm. The reader should refer to Appendix A of \cite{lovelace_binary-black-hole_2008} and Ref.~\cite{Owen:2017yaj} for more background and references.

There is no unambiguous way to define the spin of a dynamical black hole horizon in a BBH system. We usually compute a quasilocal spin definition that reduces to the expected well-defined value in the limit of infinite separation. AKV spin is one of these quasilocal spin definitions, constructed using approximate Killing vectors of the black hole horizons.

We start by setting up the conventions used throughout the paper. We
work in the standard 3+1 decomposition
\cite{NR_Baumgarte,NR_shibata,NR_Alcubierre}, with the spatial slice
denoted by $\Sigma$. The spatial metric, extrinsic curvature, and the
two-dimensional apparent horizon are denoted by $g_{ij}$, $K_{ij}$,
and $\mathcal{H}$, respectively. Lower-case Latin letters index the
spatial tensors, and upper-case Latin letters index the tensors on the
two-dimensional horizon surface. We also put a circle above the geometric quantities related to $\mathcal{H}$ to separate them from the geometric quantities on the
spatial slice $\Sigma$.

In NR, the standard definition \cite{lovelace08_73} of spin angular momentum is given by 
\begin{align}
S=\frac{1}{8 \pi} \oint_{\mathcal{H}} \phi^i s^j K_{i j} d A
\label{eqn:spin angular momentum}
\end{align}
where  $s^j$ is the outgoing unit normal to $\mathcal{H}$. Here $\vec{\phi}$ is an ``azimuthal'' vector field, tangent to $\mathcal{H}$, which carries the information about the spin axis.

It can be shown that Eq.(\ref{eqn:spin angular momentum}) is conserved \cite{lovelace08_73} when $\vec{\phi}$ is a Killing vector of the dynamical horizon world tube. For an isolated Kerr black hole, this Killing vector defines the spin axis of the horizon, but there is no reason why such a Killing vector field should exist for the horizon of a black hole in a binary system. The next best thing one can use is the vector field that is closest to being a Killing vector field; in other words, an approximate Killing vector. We now describe the method used in SpEC to construct these approximate Killing vector fields. We start from Killing's equation for a vector field $\phi_{B}$
\begin{align}
D_{(A} \phi_{B)}=0,
\end{align}
where $D$ is the covariant derivative compatible with the metric $\mathring{g}_{A B}$ on the two-dimensional horizon surface.
This condition can be split into two conditions: first, that the field has no divergence
\begin{align}
\Theta:= \mathring{g}^{A B} D_A \phi_B=0,
\end{align}
where $\mathring{g}_{A B}$ is the metric on the horizon surface.
The second condition is that the field has no shear

\begin{align}
\sigma_{A B}:=D_{(A} \phi_{B)}-\frac{1}{2} \mathring{g}_{A B} \Theta=0.
\label{eqn:shear in terms of phi}
\end{align}

If $\vec{\phi}$ is not a Killing vector field, we cannot satisfy both conditions simultaneously, so we will choose a $\vec{\phi}$ that is divergence-free and has the smallest possible shear. Any divergence-free vector field tangent to the apparent horizon can be written as
\begin{align}
    \phi^{A} = \epsilon^{AB}  D_{B} z,
    \label{eqn:phi in terms of z}
\end{align}
where $z$ is some smooth function and $\epsilon^{AB}$ is the Levi-Civita tensor. The problem of finding the approximate Killing vector then reduces to finding the $z$ that gives the smallest shear, or equivalently, minimizes the norm $\|\sigma\|^2$:
\begin{align}
\|\sigma\|^2:=\oint_{\mathcal{H}} \sigma_{B C} \sigma^{B C} d A .
\end{align}

Using Eqs.(\ref{eqn:shear in terms of phi}) and (\ref{eqn:phi in terms of z}) and integrating by parts, we get
\begin{align}
\|\sigma\|^2=\oint_{\mathcal{H}} z H z d A.
\label{eqn:shear norm in terms of H and z}
\end{align}
Here $H$ is
\begin{align}
H z=D^4 z+\mathring{R} D^2 z+(D^A \mathring{R})(D_A z)
\label{eqn:equations for H in terms of laplacians}
\end{align}
where $D^2$ is the Laplacian on the apparent horizon and $\mathring{R}$ is the Ricci scalar. If we minimize Eq.(\ref{eqn:shear norm in terms of H and z}) with respect to $z$, we obtain the trivial solution where $z$ is constant and $\vec{\phi}$ is zero. To avoid this, we will require that the minimized $\vec{\phi}$ has a positive norm,
\begin{align}
\oint_{\mathcal{H}} \phi^A \phi_A d A > 0
\label{eqn:Norm condition to be enforced using Lagrange multiplier}
\end{align}
Using a Lagrange multiplier $\lambda$ and the fact that $\phi^A \phi_A = D^A z D_A z$, we get
\begin{align}
I[z]:=\oint_{\mathcal{H}} z H z d A+\lambda\left(\oint_{\mathcal{H}} D^A z D_A z d A-\mathcal{N}\right),
\label{eqn:Optimizer with the lagrange multiplier and N}
\end{align}
where $\mathcal{N}$ is a still-undetermined positive constant. Fixing the value of $\mathcal{N}$ is equivalent to normalizing the approximate Killing vector field, which determines both the spin axis and the spin magnitude.

Minimizing the functional (\ref{eqn:Optimizer with the lagrange multiplier and N}) with respect to $z$ gives a generalized eigenvalue problem:
\begin{align}
  H z = \lambda D^2 z
  \label{eqn:generalized eigenvalue problem}
\end{align}
We can try to discretize Eq.(\ref{eqn:generalized eigenvalue problem}) and solve it numerically. The only remaining issue is that both $H$ and $D^2$ still share a kernel in which $z$ is constant, making this eigenvalue problem singular. The standard way to deal with such issues is to remove the common kernel space and then solve for $z$. In our case, this is trivial because we are discretizing using spherical harmonics and can just remove the $Y_{00}$ mode.

On a metrically round two-sphere, the scalar potentials associated with rotational Killing vector fields are the three $\ell=1$ spherical harmonics. After removing the constant $Y_{00}$ mode, we therefore retain the three least-shear eigenvectors, which provide a natural generalization of this three-dimensional rotation subspace to a deformed horizon. Although only the first eigenvector strictly minimizes the shear, the three least-shear eigenvectors together define the preferred subspace of generalized rotation generators.

From Eq.~(\ref{eqn:shear norm in terms of H and z}) and Eq.~(\ref{eqn:generalized eigenvalue problem}), we can see that the amount of shear is proportional to the magnitude of $\lambda$. Therefore, once we have discretized Eq.~(\ref{eqn:generalized eigenvalue problem}), the next step is to find the three smallest-magnitude eigenvalues and their corresponding eigenfunctions. We can use these eigenfunctions $z$ to compute the components of the spin along the corresponding axes. However, we still need to determine the spin magnitude by normalizing the eigenfunctions, which is equivalent to fixing $\mathcal{N}$ in Eq.~(\ref{eqn:Optimizer with the lagrange multiplier and N}). To fix this normalization, we use

\begin{align}
\oint_{\mathcal{H}}(z-\langle\langle z\rangle\rangle)^2 d A=\frac{A^3}{48 \pi^2}
\label{eqn:normalization condition for z}
\end{align}
where $\langle\langle z\rangle\rangle$ is the average of $z$ over the surface and $A$ is the surface area of the apparent horizon. We use this equation to normalize the three eigenfunctions, after which their Euclidean norms give the spin magnitude. Mathematically, Eq.~(\ref{eqn:normalization condition for z}) holds only for a Kerr metric, where $z$ is a true rotation generator of the horizon. Therefore, when using it for normalization, we assume that the evolving black holes are approximately Kerr.

Using the AKV potential $z$ and the normalization condition in Eq.(\ref{eqn:normalization condition for z}) is the most robust way we know to get the spin magnitude, and it is what we use by default in SpEC. Determining the spin axis is much more complicated, and not all definitions are based on $z$. Various definitions of the spin axis, along with their trade-offs, are discussed in \cite{Owen:2017yaj}.

Now that we have some understanding of why we need to solve a generalized eigenvalue problem, we next describe the current algorithm and explain why it becomes problematic for complex systems.

\section{\label{sec:Current AKV algorithm} Current AKV algorithm}

Call the matrices formed by discretizing the operators $H$ and $D^2$ in Eq.(\ref{eqn:generalized eigenvalue problem}) $M$ and $B$, respectively. Now we have a generalized eigenvalue problem $Mx = \lambda Bx$, and the obvious way to obtain the eigenvectors is to use a standard library such as LAPACK \cite{lapack_users_guide}. To understand why this becomes an issue, we need to understand how big these matrices are and how they grow with an increase in the complexity of the apparent horizons.

Apparent horizons in SpEC are represented using spherical harmonics. When the black holes are far away from each other, their apparent horizons are mostly spherical and can be approximated using a small number of spherical harmonics. On the other hand, when the black holes are about to merge, the horizons become very distorted, and we need a larger number of spherical harmonics to approximate them. We will denote the highest spherical harmonic used in representing the apparent horizon as $L$. For the simulations that we have done so far, $L$ can be as small as 15 during the inspiral, but in some complex cases it can get as large as 90 near the merger. As we push to evolve more and more complex systems at higher and higher resolution, we expect the largest required value of $L$ to increase even further.

\begin{figure}[t]
  \centering
  \includegraphics[width=0.75\textwidth]{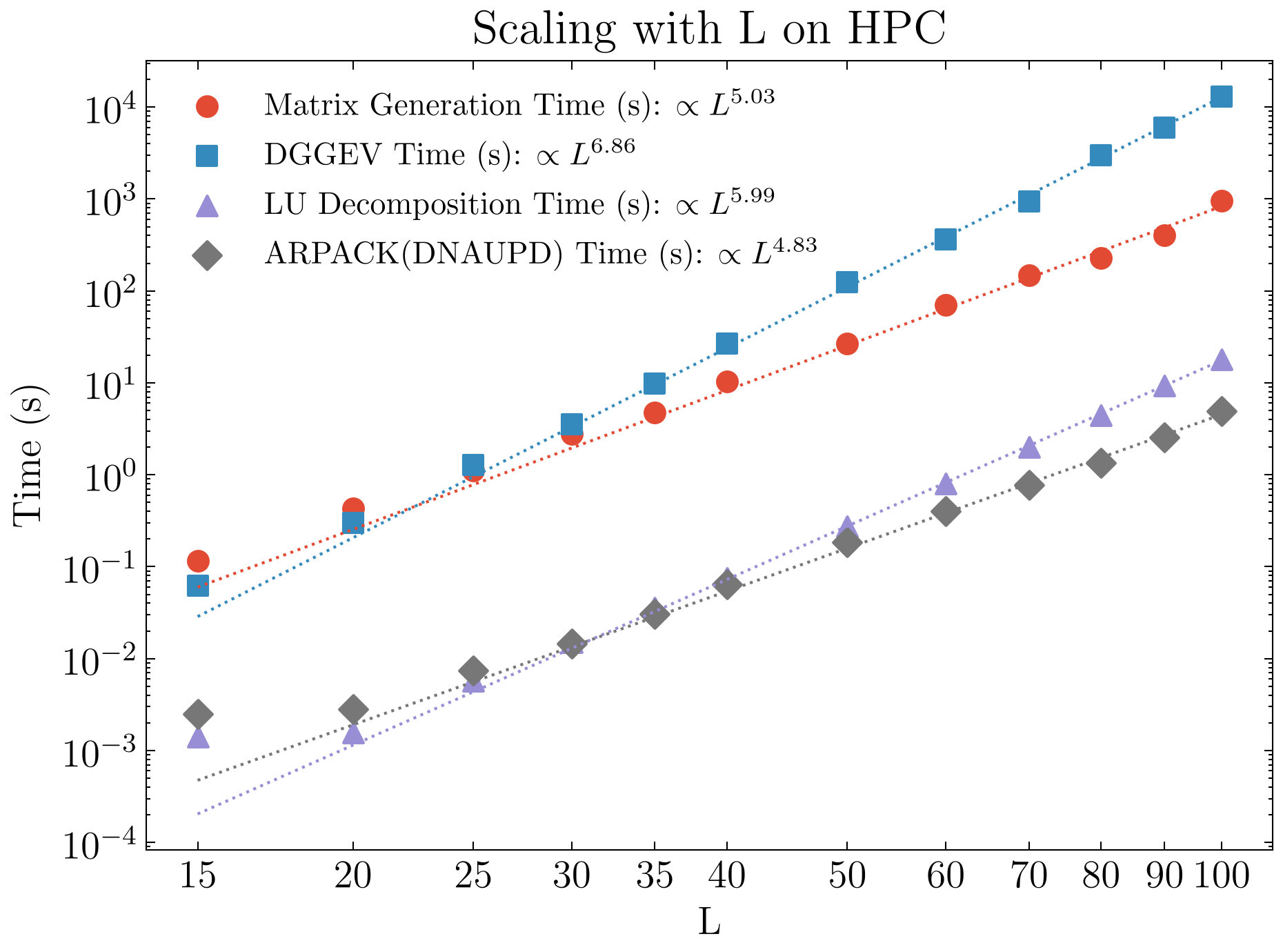}
  \caption{Scaling of various parts of the algorithm with the highest spherical harmonic mode $L$. The runs were performed on Caltech HPC (Intel(R) Xeon(R) Platinum 8276 CPU @ 2.20GHz). The solid dots are the real data, while the dotted lines show the best fits. The fits were weighted such that the measurements at higher $L$ have more weight, as they are less noisy. The legends include the scaling from these fits. We see that the scalings of matrix generation and \texttt{dggev} are slightly worse than the theoretical expectations of $\mathcal{O}(L^4)$ and $\mathcal{O}(L^6)$, respectively. Note that the new AKV algorithm using ARPACK + LU decomposition (purple + black lines) is orders of magnitude faster than the current AKV algorithm using \texttt{dggev} (blue line). By a large margin, the current bottleneck is matrix generation (red line).}
  \label{fig:HPC_all_in_one}
\end{figure}

The number of rows in the $M$ and $B$ matrices is $N = (L-1)^2 -1$. The reason for $L-1$ is that even though we have $L+1$ modes, we take second derivatives, and that pollutes the two highest modes. The term $-1$ is because we ignore $Y_{00}$, as described in Sec.(\ref{sec:AKV spin definition}). We only need the eigenvectors corresponding to the three smallest-magnitude eigenvalues, but using LAPACK's dense solver (\texttt{dggev}) means we always compute all eigenvalues ($L^2$ in number)\footnote{There are other LAPACK routines that can select the three required eigenvalues and compute only their eigenvectors, but they provide very little speedup \cite[Sec.~3.4.3]{lapack_users_guide}. Before finding the eigenvectors, LAPACK performs an eigensystem reduction, which is the most costly operation. Once the reduction is done, finding the eigenvectors themselves is comparatively cheap, which means that finding just a few eigenvectors is not significantly faster than finding all of them.}. The theoretical time complexity of LAPACK's \texttt{dggev} solver is $\mathcal{O}(N^3)$, or equivalently $\mathcal{O}(L^6)$. This is the reason why the cost of \texttt{dggev} is irrelevant for most of the inspiral until we get close to merger, where it starts to increase rapidly. In our simulations, we have found the scaling to be closer to $\mathcal{O}(L^7)$, as shown in Figs. (\ref{fig:HPC_all_in_one}) and (\ref{fig:PC_all_in_one}). We suspect that the slightly worse scaling may be due to memory pressure since the matrix has $\mathcal{O}(L^4)$ elements, although we have not rigorously verified this.

\begin{table}[h]
\centering
\begin{tabular}{c c c c c}
\toprule
$L$ & Mat gen (s) & LU (s) & ARPACK (s) & \texttt{dggev} (s) \\
\midrule
15 & 0.117 & 0.00244 & 0.00419 & 0.0734 \\
20 & 0.428 & 0.00158 & 0.00286 & 0.301 \\
25 & 1.12 & 0.00594 & 0.00788 & 1.29 \\
30 & 2.77 & 0.015 & 0.0145 & 3.62 \\
35 & 4.76 & 0.0357 & 0.0309 & 9.95 \\
40 & 10.3 & 0.0754 & 0.0646 & 26.4 \\
50 & 26.7 & 0.276 & 0.183 & 125 \\
60 & 71 & 0.806 & 0.404 & 361 \\
70 & 147 & 2.01 & 0.775 & 949 \\
80 & 228 & 4.47 & 1.35 & 3.02e+03 \\
90 & 407 & 9.01 & 2.06 & 5.95e+03 \\
100 & 965 & 18 & 4.7 & 1.31e+04 \\
\bottomrule
\end{tabular}
\caption{Timing data for matrix generation, LU decomposition, ARPACK iterations, and LAPACK's \texttt{dggev} on Caltech HPC (Intel(R) Xeon(R) Platinum 8276 CPU @ 2.20GHz). The column showing the ARPACK timing also includes the time taken by the LU solves required for each iteration.}
\label{table:timing_data_HPC}
\end{table}

From Table~(\ref{table:timing_data_HPC}), we see that the runtime cost of \texttt{dggev} goes from $0.07$ s to $13{,}100$ s as $L$ goes from 15 to 100.
For some simulations, each such spin calculation took over 30 minutes on a single core, tying up that core while the rest of the cores wasted resources waiting for it to finish. This waste of computing resources would only get worse with the need to run more accurate simulations. In the next section, we outline a new algorithm that uses the iterative eigensolver ARPACK to efficiently compute the three required eigenvectors, providing a large speedup.

\section{\label{sec:Sparse AKV algorithm} ARPACK-based AKV algorithm}

From the previous Section~\ref{sec:Current AKV algorithm}, we saw that the primary cause of the poor scaling of the current AKV algorithm is the LAPACK routine \texttt{dggev}, which solves for the whole eigensystem. To find the AKV spin, we only require the three smallest-magnitude eigenvalues and their corresponding eigenvectors. There are standard iterative algorithms, implemented in libraries like ARPACK \cite{arpack_guide}, that can find just the required eigenpairs very efficiently. For a highly deformed horizon, where $L\approx100$, \texttt{dggev} finds all $\approx 10^4$ eigenpairs, and we can significantly reduce the required computation by finding only the three required eigenpairs using ARPACK. To find an eigenpair, ARPACK just needs a series of matrix-vector products until it converges to the required solution. Because we only need three eigenpairs, the theoretical computational complexity of using ARPACK is $\mathcal{O}(L^4)$, which is much better than that of \texttt{dggev} ($\mathcal{O}(L^6)$).

In the following sections, we will first discuss issues that ARPACK has in trying to find the eigenvalues with very small magnitude and how to resolve them. Then we will discuss the cost of assembling the matrices $M$ and $B$. Finally, we will discuss the cost of getting the action of the inverse of these matrices required by ARPACK.

\subsection{\label{ssec:Shift Invert transform} Shift-invert transform}

The number of iterations needed by ARPACK to converge depends on the distribution of the eigenvalues. If we are trying to find the first few largest-magnitude eigenvalues and they are well separated, then the convergence is very fast. On the other hand, if we are trying to find the smallest-magnitude eigenvalues, convergence is very slow, and it might not even converge for some systems. For problems like ours, where we need to find the smallest-magnitude eigenvalues, the way to maintain good convergence is to use the shift-invert transform.

The shift-invert transform is a method for transforming our eigenvalue problem into another in which the desired eigenvalues become large in magnitude and well separated from their neighbors. The original problem is
\begin{equation}
    M x = \lambda B x
\end{equation}
After applying the shift-invert transform, it becomes:
\begin{equation}
    (M - \sigma B)^{-1} B x = \nu x , 
    \label{eqn:shift_inverted_equation}
\end{equation}  
where $\sigma$ is a free parameter we can choose, and
\begin{equation}
    \nu = \frac{1}{\lambda - \sigma}.
    \label{eqn:shift_invert_evals_relation}
\end{equation}

The first thing to observe is that both problems have the same eigenvectors, while the corresponding eigenvalues are related by Eq.(\ref{eqn:shift_invert_evals_relation}). From Eq.(\ref{eqn:shift_invert_evals_relation}), we see that among the new eigenvalues in Eq.~(\ref{eqn:shift_inverted_equation}), $|\nu|$ is largest for the values of $\lambda$ that are closest to $\sigma$. So if we want to find the three smallest-magnitude eigenvalues, we should just set $\sigma = 0$ and ask ARPACK for the largest-magnitude transformed eigenvalues. This will work until one of the eigenvalues gets too close to $\sigma$, and then the condition number of the matrix $M-\sigma B$ will diverge. One could consider shifting the value of $\sigma$ slightly away from zero, but then we would find the three eigenvalues closest to $\sigma$, not the three smallest-magnitude eigenvalues.

\begin{figure}[t]
  \centering
  \begin{subfigure}[t]{0.48\textwidth}
    \centering
    \includegraphics[width=\linewidth]{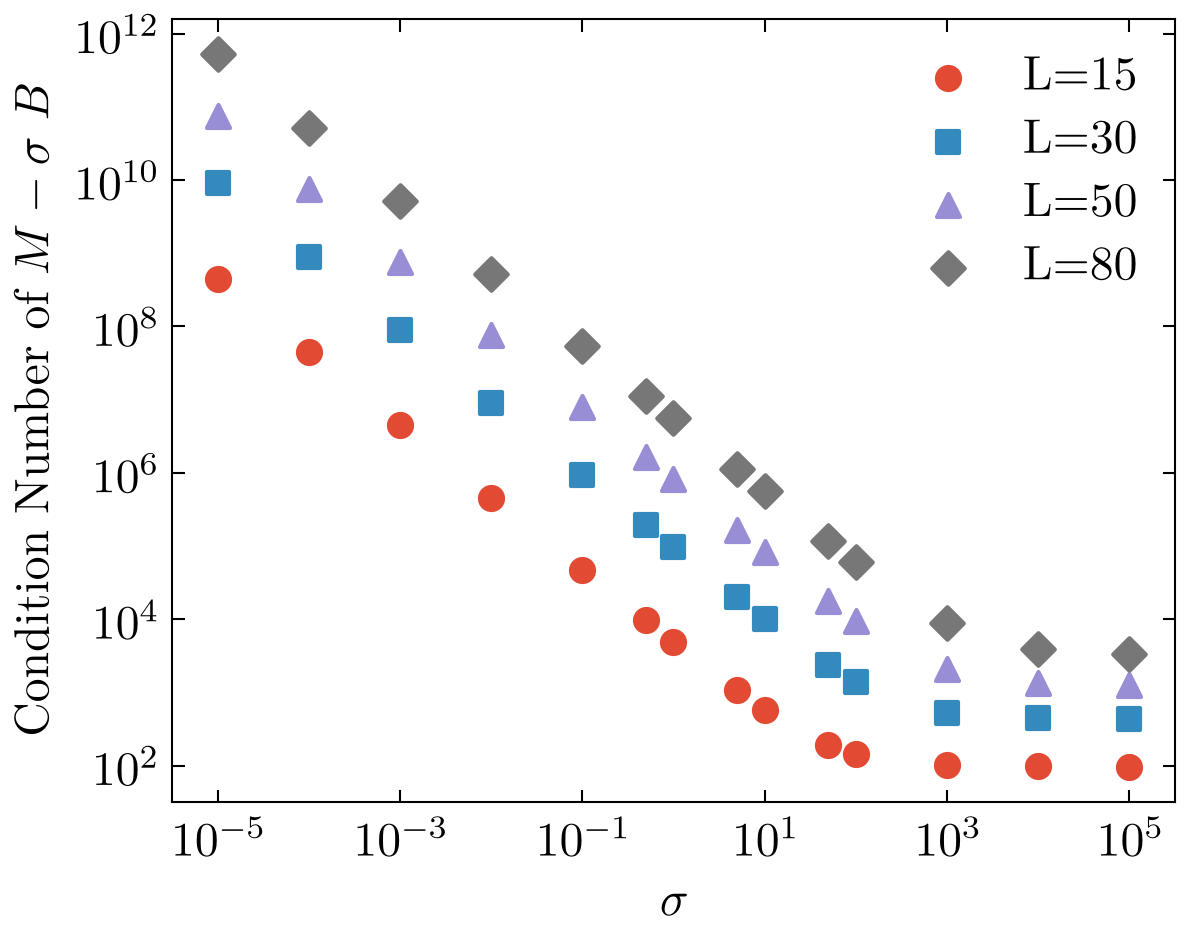}
    \caption{Condition number of $M-\sigma B$ as a function of $\sigma$ for several values of $L$. The conditioning worsens as $L$ increases.}
    \label{fig:HPC_condition_number}
  \end{subfigure}\hfill
  \begin{subfigure}[t]{0.48\textwidth}
    \centering
    \includegraphics[width=\linewidth]{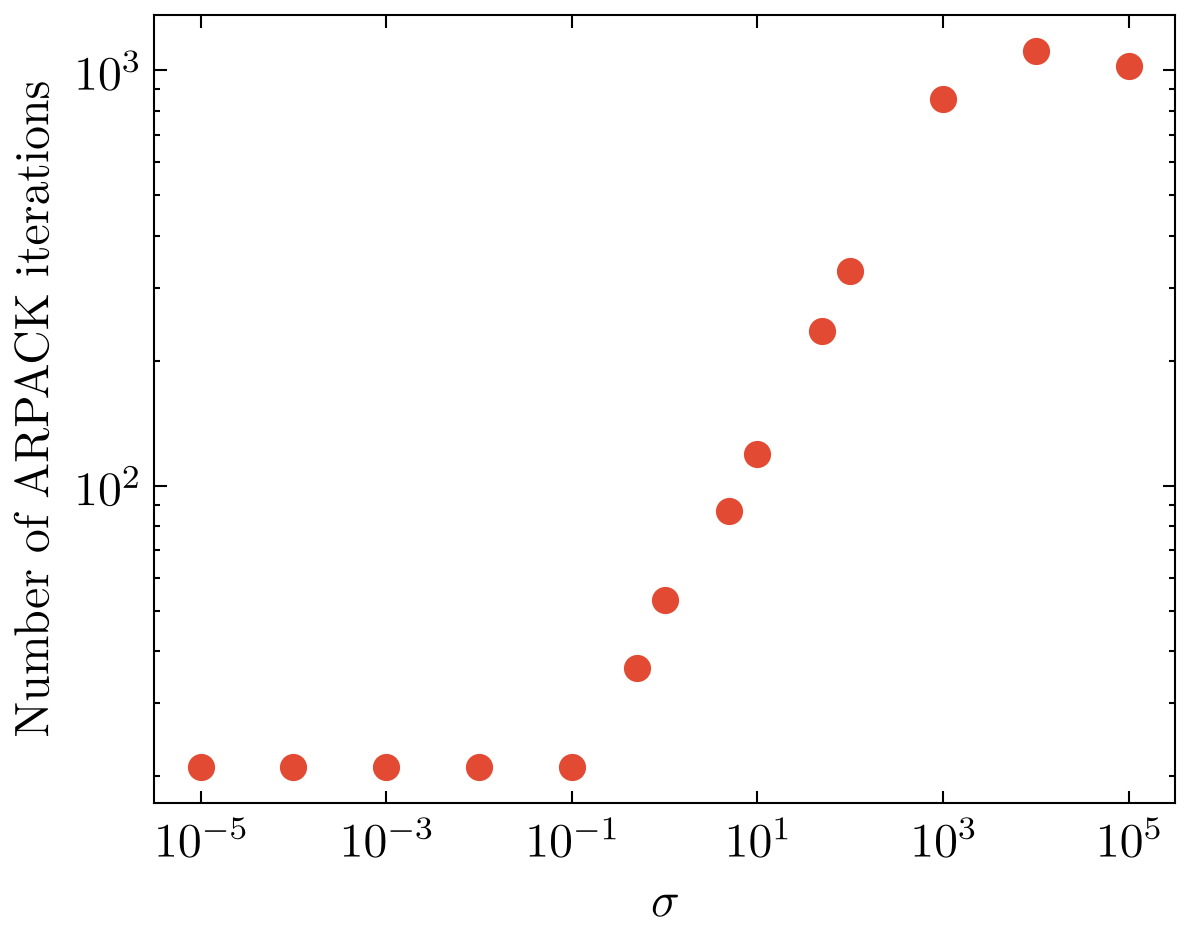}
    \caption{Number of ARPACK iterations required for convergence at $L=50$. Each ARPACK iteration requires one LU solve.}
    \label{fig:LU_solve_nums_L50}
  \end{subfigure}
  \caption{Trade-off in choosing the shift $\sigma$. Both panels use matrices generated from a Kerr-Schild horizon with a mass of 1 and spin $(0,0,0.5)$. Small values of $\sigma$ make $M-\sigma B$ poorly conditioned, whereas large values increase the number of ARPACK iterations. Although the quantitative behavior depends on the horizon geometry, the same general trend appears in the systems we tested. We use $\sigma=0.1$, which provides a good balance between conditioning and convergence.}
  \label{fig:sigma_dependence}
\end{figure}

This is not an issue if we are finding these eigenvectors in isolation, because we can always adjust the value of $\sigma$ slightly and then try again. In a long-running simulation where the eigenvalues keep changing, we cannot afford to suddenly have slightly wrong eigenvectors because one of the eigenvalues drifts too close to the value of $\sigma$. Fortunately, since we are using the norm described in Eq.(\ref{eqn:Norm condition to be enforced using Lagrange multiplier}), the matrices $M$ and $B$ are positive- and negative-semidefinite, respectively, which implies that the generalized eigenvalues $\lambda$ are non-positive.
Thus, any positive value of $\sigma$ will always give us the correct results. If we choose a small value of $\sigma$, the eigenvalues we need will be well separated and large in magnitude, making ARPACK very efficient at finding them. But we do not want $\sigma$ to be too small because that makes the matrix $M-\sigma B$ ill-conditioned. This means that there is an ideal value of $\sigma$ that makes the magnitude of the transformed eigenvalues corresponding to the smallest-magnitude eigenvalues as large as possible, without making the matrix $M - \sigma B$ too ill-conditioned. Figs.(\ref{fig:HPC_condition_number}) and (\ref{fig:LU_solve_nums_L50}) show how the condition number of $(M- \sigma B)$ and the number of iterations taken by ARPACK depend on $\sigma$. For these two figures, $M$ and $B$ were generated using the horizon of a Kerr-Schild metric with spin $(0,0,0.5)$ and a mass of 1. These figures change for different shapes of the apparent horizons, but the same overall trend is always present. 

For the systems that we tested, the eigenvalues found by ARPACK were very accurate even for very small values of $\sigma$ for which the condition number became as large as $10^{12}$. In fact, for large values of $L$, the eigenvalues found by ARPACK were more accurate than those found by \texttt{dggev}, especially when one of the eigenvalues was very close to zero. In practice, the accuracy of the eigenvalues found by either solver is adequate for our purposes because we are limited by errors in other parts of the algorithm, such as finding the apparent horizons. For now, we have decided to choose $\sigma = 0.1$ as the default value for our simulations, which allows for rapid convergence without generating very ill-conditioned matrices. Note that not all norms lead to eigenvalue problems in which all eigenvalues are negative. For example, the norm used by Cook and Whiting \cite{lovelace08_41} will not have this property, making it hard to choose a stable and effective value for $\sigma$.

\begin{figure}[h]
\centering
\begin{tikzpicture}[
    node distance=1.6cm,
    block/.style={draw, rectangle, rounded corners, minimum width=4.5cm, minimum height=1.1cm, align=center, thick},
    input/.style={draw, ellipse, minimum width=2.2cm, minimum height=1.0cm, align=center, thick},
    arrow/.style={->, thick},
    dashedbox/.style={draw, dashed, rounded corners, inner sep=0.3cm, thick}
]

\node[block, fill=orange!20] (build) {Assemble matrices $M$ and $B$};
\node[block, fill=orange!20, below of=build] (factor) {Form $M-\sigma B$, compute its \\ LU decomposition, and cache it};

\node[input, fill=blue!10, below=1.8cm of factor] (x) {$x$ from ARPACK};
\node[block, fill=green!20, below of=x] (solve) {Solve $(M-\sigma B)y=Bx$ \\ using the LU factors};
\node[input, fill=blue!10, below of=solve] (y) {$y$ to ARPACK};

\draw[arrow] (build) -- (factor);
\draw[arrow] (factor) -- (x);

\draw[arrow] (x) -- (solve);
\draw[arrow] (solve) -- (y);

\draw[arrow] (y.east) .. controls +(2.0,0) and +(2.0,0) .. 
    node[midway, right] {} (x.east);

\node[
    dashedbox,
    fit=(x)(solve)(y),
    label={[rotate=90, anchor=south]left:{\textbf{ARPACK iteration loop}}}
] {};

\end{tikzpicture}
\caption{Steps in the new AKV algorithm using LU decomposition and ARPACK. We first do the one-time setup, which requires assembling the matrices $M$ and $B$ and computing the LU decomposition of $(M-\sigma B)$. ARPACK needs the action of $(M-\sigma B)^{-1}B$ on the vector $x$ that it provides. To construct this vector, we solve $(M-\sigma B)y = Bx$ for $y$ using the LU factors.}
\label{fig:ARPACK_call_graph_shift_invert_generalized}
\end{figure}
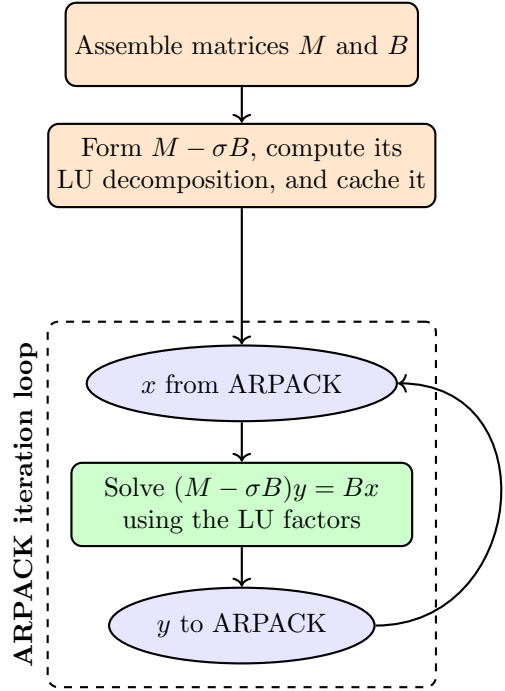

\subsection{\label{ssec:Cost of assembling the matrices} Cost of assembling the matrices}

Recall that matrices $M$ and $B$ represent the discretization of the operators $H$ and $D^2$ that appear in Eq.(\ref{eqn:generalized eigenvalue problem}). It is not straightforward to derive explicit formulas for $M$ and $B$, but SpEC has two functions, $f_M(x)$ and $f_B(x)$, that give the action of the matrices $M$ and $B$ on a vector $x$. To generate the matrices, we need to operate on $N$ Cartesian basis vectors, which can be costly. 

It is important to consider the cost of the matrix generation because, in theory, we do not even need to construct these matrices when using ARPACK. From Table (\ref{table:timing_data_HPC}), we see that the cost of constructing the matrix exceeds that of \texttt{dggev} for $L<25$. We expect the time complexity of generating the matrices $M$ and $B$ to be around $\mathcal{O}(N^2)$: one factor of $N$ arises because we need to act on $N$ basis vectors, and another because the cost of the functions $f_M(x)$ and $f_B(x)$ is $\mathcal{O}(N)$. From Figs. (\ref{fig:HPC_all_in_one}) and (\ref{fig:PC_all_in_one}), we see that the cost lies between $\mathcal{O}(L^{4.5})$ and $\mathcal{O}(L^5)$ depending on the system we are running on. This scaling behavior means that the cost of \texttt{dggev} quickly overtakes the cost of constructing the matrices, becoming the main bottleneck.

Like many iterative algorithms, ARPACK can work directly with the functions $f_M(x)$ and $f_B(x)$, saving both memory and computation. Working in this matrix-free manner was our original plan, but it failed for reasons described in detail in Appendix (\ref{sec:Matrix-free version of the AKV algorithm}). The final version we ended up using involves constructing the full matrices $M$ and $B$, and we will assume that we have access to those in the later sections.

\subsection{\label{ssec:action-of-shift-invert} Cost of \texorpdfstring{$(M-\sigma B)^{-1}$}{(M-sigma B)^(-1)}}

When ARPACK solves a sparse eigenvalue problem, for example $Mx=\lambda x$, it does so iteratively. On each iteration, ARPACK provides a trial eigenvector $x$ to the user, the user code computes $Mx$ and sends it to ARPACK, and then ARPACK updates its trial eigenvector for the next iteration. Similarly, for the shift-invert system, ARPACK still provides a trial eigenvector $x$ to the user. The user code computes $(M-\sigma B)^{-1}Bx$ and sends it to ARPACK, which updates $x$ until convergence.
A typical iteration of ARPACK solving a shift-invert system is shown in Fig.(\ref{fig:ARPACK_call_graph_shift_invert_generalized}). Note that it requires us to solve the linear system $(M-\sigma B) y=Bx$. For $\sigma=0.1$, ARPACK usually needs around 30 iterations to get the three smallest-magnitude eigenvalues and their corresponding eigenvectors. Of course, we do not want to take the inverse of the matrix explicitly, so two options are solving the linear system using LU decomposition or using iterative methods. For reasons mentioned in Appendix \ref{sec:Matrix-free version of the AKV algorithm}, we will avoid iterative methods and focus on using LU decomposition to solve the linear system.

Now the theoretical issue with using LU decomposition is that it scales as $\mathcal{O}(N^3) \approx \mathcal{O}(L^6)$, which is the same as \texttt{dggev}. This means that scaling-wise, we have just pushed the problem from \texttt{dggev} to the LU solve. But unlike \texttt{dggev}, LU factorization is much more optimized\footnote{Part of the reason is that LU decomposition makes very good use of BLAS Level 3 routines while \texttt{dggev} does not. In general, the number of flops required for an LU decomposition is much smaller than that required by \texttt{dggev}. Thus, in practice, LU decomposition is much faster, even if both theoretically scale as $\mathcal{O}(N^3)$.} and runs at least two orders of magnitude faster than \texttt{dggev}, as we can see from Table \ref{table:timing_data_HPC}.

Of course, because of the $\mathcal{O}(L^6)$ scaling for very large $L$, the
cost of LU decomposition will eventually overtake the cost of matrix
generation. Extrapolating the measured scaling, for the systems we are
solving this will happen only when $L \approx 10^3$, well beyond our
expected use cases.

Note that while ARPACK requires the actions of $(M- \sigma B)^{-1}$
for each iteration it takes, we only need to do the costly LU
factorization once and then cache it. Getting the action of $(M-
\sigma B)^{-1}$ on a vector from a cached LU factorization is almost as cheap
as a simple matrix-vector product. From Table
\ref{table:timing_data_HPC}, we can see that the cost of doing the LU
decomposition is negligible compared to the cost of constructing the
matrices.

\section{\label{sec:Future improvements} Future improvements}

The first possible improvement to try is to get the matrix-free version of the algorithm working. The idea would be to generate the full matrices every 100th step or so to use as a preconditioner for the iterations. If the spin does not change very rapidly, then the matrix should also change slowly, and the same preconditioner could be used for multiple time steps. The main issue is that, close to the merger, spins change rapidly, and we may end up calculating the full matrices much more frequently. 

Another possibility is to obtain a better preconditioner using the analytical properties of the matrices, which will allow us to work in a matrix-free manner, avoiding the cost of constructing the matrices $M$ and $B$. The construction of the matrices requires $L^2$ calls to the functions $f_M$ and $f_B$, which is the bottleneck in the new algorithm. When using a matrix-free version, we need to use an iterative linear solver like GMRES, which will call these functions multiple times to get the action of $(M-\sigma B)^{-1}$. In our testing, we have found that with $\sigma = 0.1$, ARPACK typically requires fewer than 30 iterations to converge. For the matrix-free version to be faster, the total number of calls to $f_M$ and $f_B$ made by the ARPACK and linear-solver iterations needs to stay below the $N\sim L^2$ calls required to generate the full matrices, which is very hard to do without a good preconditioner. 

Note that matrix generation is an embarrassingly parallel problem in which we apply the functions $f_{M}$ and $f_{B}$ to unit vectors one by one to construct the matrices. For large values of $L$, we can gain a good speedup by doing this in parallel, especially for next-generation codes like SpECTRE \cite{spectrecode} with better parallelism infrastructure than SpEC.

\section{\label{sec:Conclusion} Conclusion}

We have shown that the new algorithm based on the eigensystem solver (ARPACK) and LU decomposition is close to two orders of magnitude faster than the original one based on the dense eigensystem solver (LAPACK). If we include the time required to construct the matrices, which is required by both algorithms, then the speedup is closer to one order of magnitude. We tried to use a fully matrix-free algorithm, but were unable to find preconditioners for which the overall cost of the iterations and the preconditioner construction was lower than the cost of constructing the full matrix. However, the idea of using the fixed full matrix for a number of timesteps as a preconditioner is a promising future improvement to explore.

\appendix
\section{\label{app:Timing data PC} PC timing data}

\begin{figure}[t]
  \centering
  \includegraphics[width=0.75\textwidth]{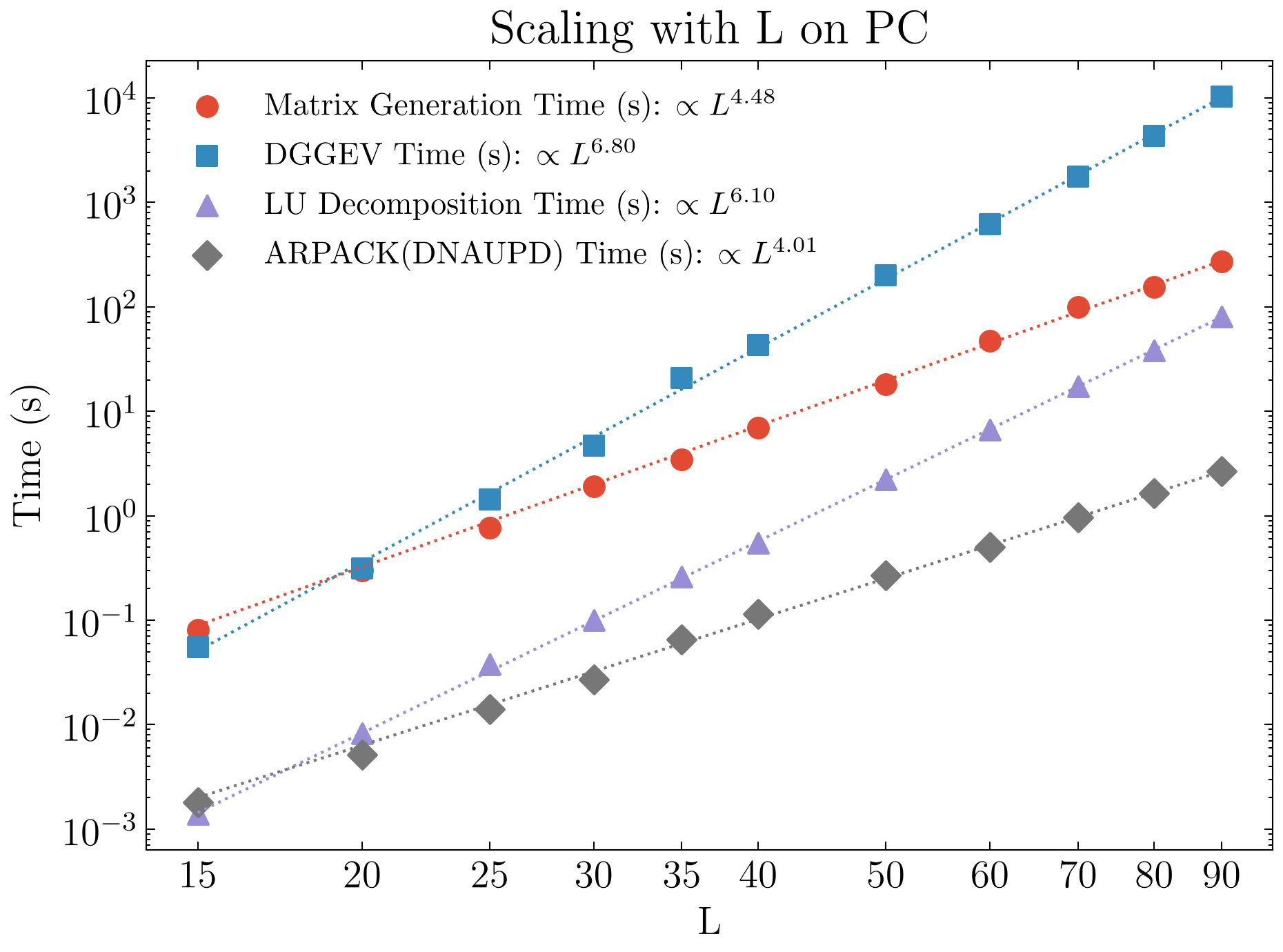}
  \caption{Scaling of various parts of the algorithm with the highest spherical harmonic mode $L$. The runs were performed on a consumer-grade processor (Intel(R) Core(TM) i9-10900X CPU @ 3.70GHz). Labels and conclusions are the same as for Fig.(\ref{fig:HPC_all_in_one}). Note that the LU decomposition is $\approx 10 \times\text{slower}$ while the matrix generation is $\approx 2 \times\text{faster}$ when compared to running on an HPC system, as shown in Fig.(\ref{fig:HPC_all_in_one}).}
  \label{fig:PC_all_in_one}
\end{figure}

In this section, we show the same data as in Table \ref{table:timing_data_HPC} and Fig. \ref{fig:HPC_all_in_one}, but generated on a PC (Intel(R) Core(TM) i9-10900X CPU @ 3.70GHz). This was done to get an idea of how things scale with differences in libraries, memory bandwidth, etc. The first thing to note is that matrix generation is around twice as fast now, while LAPACK operations are slower (LU is around 10 times slower and \texttt{dggev} is around 2 times slower). The LAPACK operations being slower is expected because HPC is using vendor-optimized libraries and has access to more SIMD lanes. The matrix generation being faster on a local PC is not too surprising either, because for single-core jobs, desktop CPUs are faster than cluster ones and have access to more memory bandwidth per core.

The primary conclusion remains the same, though. The ARPACK + LU factorization path is much faster than the LAPACK (\texttt{dggev}) path, and the main bottleneck now is the matrix generation. What changes is the value of $L$ at which the LU decomposition becomes the bottleneck. Extrapolating the measured scaling, this goes from $L\approx 10^3$ to $L \approx 200$. For all practical purposes, $L \approx 200$ is still too big. For all current HPC systems, we will be in the regime shown in Fig.(\ref{fig:HPC_all_in_one}), where the matrix generation remains the bottleneck for much larger values of $L$.

\begin{table}[h]
\centering
\begin{tabular}{c c c c c}
\toprule
$L$ & Mat gen (s) & LU (s) & ARPACK (s) & \texttt{dggev} (s) \\
\midrule
15 & 0.0809 & 0.00139 & 0.00177 & 0.0464 \\
20 & 0.304 & 0.00827 & 0.0053 & 0.306 \\
25 & 0.761 & 0.0319 & 0.0125 & 1.44 \\
30 & 1.93 & 0.101 & 0.0275 & 5.05 \\
35 & 3.46 & 0.273 & 0.0686 & 21.8 \\
40 & 6.97 & 0.545 & 0.108 & 45.4 \\
50 & 18.2 & 2.18 & 0.269 & 200 \\
60 & 47.5 & 6.7 & 0.503 & 615 \\
70 & 99.9 & 17.2 & 0.954 & 1.77e+03 \\
80 & 153 & 38.4 & 1.64 & 4.34e+03 \\
90 & 274 & 79 & 2.59 & 1.03e+04 \\
\bottomrule
\end{tabular}
\caption{Timing data for matrix generation, ARPACK, LU decomposition, and LAPACK's \texttt{dggev} on PC (Intel(R) Core(TM) i9-10900X CPU @ 3.70GHz).  The column showing the ARPACK timing also includes the time taken by the LU solves required for each iteration.}
\label{table:timing_data_PC}
\end{table}

\section{\label{sec:Matrix-free version of the AKV algorithm} Matrix-free version of the AKV algorithm}

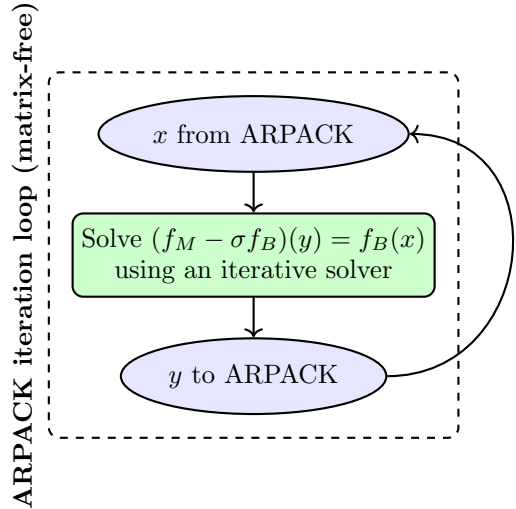
\begin{figure}[h]
\centering
\begin{tikzpicture}[
    node distance=1.6cm,
    block/.style={draw, rectangle, rounded corners, minimum width=4.5cm, minimum height=1.1cm, align=center, thick},
    input/.style={draw, ellipse, minimum width=2.2cm, minimum height=1.0cm, align=center, thick},
    arrow/.style={->, thick},
    dashedbox/.style={draw, dashed, rounded corners, inner sep=0.3cm, thick}
]

\node[input, fill=blue!10] (x) {$x$ from ARPACK};

\node[block, fill=green!20, below of=x] (solve) {Solve $(f_M - \sigma f_B)(y)=f_B(x)$ \\ using an iterative solver};

\node[input, fill=blue!10, below of=solve] (y) {$y$ to ARPACK};

\draw[arrow] (x) -- (solve);
\draw[arrow] (solve) -- (y);

\draw[arrow] (y.east) .. controls +(2.0,0) and +(2.0,0) .. 
    node[midway, right] {} (x.east);

\node[
    dashedbox,
    fit=(x)(solve)(y),
    label={[rotate=90, anchor=south]left:{\textbf{ARPACK iteration loop (matrix-free)}}}
] {};

\end{tikzpicture}
\caption{Matrix-free workflow for ARPACK in the generalized shift-invert case. Instead of explicitly forming $M$ and $B$, we use the functions $f_M$ and $f_B$, which are available in SpEC. At each iteration, ARPACK supplies a vector $x$, and we solve $(f_M - \sigma f_B)(y)=f_B(x)$ using an iterative solver. The result $y = (M-\sigma B)^{-1}Bx$ is returned to ARPACK for the next iteration.}
\label{fig:ARPACK_call_graph_shift_invert_matrix_free}
\end{figure}

As already mentioned in section \ref{ssec:Cost of assembling the matrices}, the original goal was not to construct the matrices $M$ and $B$ at all but to use the functions $f_M$ and $f_B$ directly. This is shown in Fig.(\ref{fig:ARPACK_call_graph_shift_invert_matrix_free}), where we skip the matrix construction and the LU decomposition.

We need to solve the linear problem $(f_M - \sigma f_B)y=f_B(x)$ to get the vector $y$ required by ARPACK. This can be done using iterative solvers like GMRES or BiCGSTAB. In almost all cases, these iterative algorithms require a good preconditioner to converge at an acceptable rate. This is particularly true in our case because the condition number of the matrix $(M - \sigma B)$ is fairly big for $\sigma=0.1$, and if we try to make $\sigma$ larger, then the number of iterations required by ARPACK increases significantly, as shown in Figs. \ref{fig:HPC_condition_number} and \ref{fig:LU_solve_nums_L50}. Note that we need the number of iterations required by ARPACK times the number of iterations required by the linear solver to stay below $N\sim L^2$. Otherwise, the whole algorithm will be as costly as constructing the full matrix.

We tried a few simple preconditioners, and the algorithm worked very well most of the time. However, the convergence became very slow in some edge cases when the horizon was very deformed. Since the spin computation is ideally only a small part of a very large simulation, we would prefer a stable but slightly inefficient algorithm over a mostly stable but fast one. And while constructing the matrices is costly, using LU decomposition gives ARPACK robust convergence and predictable runtime even when the condition number gets large. Thus, we decided to swallow the cost of constructing the full matrices $M$ and $B$ and then use LU factorization to efficiently apply the action of $(M- \sigma B)^{-1}$.

Using the matrix-free version in practice requires a good preconditioner that can be constructed cheaply. When there is no underlying structure and the matrices are dense, the standard methods for constructing non-trivial preconditioners from $f_{M}(x)$ are usually as costly as constructing the full matrix itself.

\section{\label{sec:Speed difference between finding the smallest and the largest eigenvalues} Speed difference between finding the smallest-magnitude and the largest-magnitude eigenvalues}

ARPACK converges much faster when we solve for largest-magnitude eigenvalues that are well separated. We can see this if we generate a diagonal matrix with entries $1,4,9,16, \ldots, 500^2$ and use ARPACK to find the eigenvalues. Finding the three smallest-magnitude eigenvalues requires 31302 matrix-vector products, while finding the three largest-magnitude eigenvalues only requires 430 matrix-vector products. This artificial example demonstrates the fact that ARPACK is more effective in finding the largest eigenvalues. If we use the shift-invert transform, which essentially transforms the smallest-magnitude eigenvalues into the largest-magnitude ones, ARPACK becomes much more efficient and requires only 38 inverse-operator applications to find the three smallest eigenvalues. This artificial example illustrates the importance of the shift-invert transform and the reduction in operator applications that can be achieved. In our case, for a black hole with spin $(0,0,0.5)$, we use around 20 times fewer operator applications for $L=15$ and around 200 times fewer operator applications for $L=50$, and the advantage of using shift-invert continues to increase as the matrix size grows.

\ack{
This work was supported in part by the Sherman
Fairchild Foundation, by NSF Grants PHY-2407742, PHY-2308615, and
OAC-2513338 at Cornell, and
NASA award 80NSSC26K0340; and by NSF Grants
PHY-2309211, PHY-2309231, and OAC-2513339, and
NASA award 80NSSC26K0340 at Caltech.
Computations were performed at Caltech using the Resnick HPC Center.
Figures were produced using Matplotlib and SciencePlots
\cite{matplotlib_2007,SciencePlots}.
}

\clearpage
\bibliographystyle{unsrt}
\bibliography{main,aliases}

\end{document}